\documentclass[final,3p,twocolumn]{elsarticle}

\usepackage{color}
\usepackage{graphicx}
\usepackage{amssymb}

\journal{Comput. Phys. Commun. (CCP 2010 special issue)}

\begin{document}

\begin{frontmatter}

\title{Monte Carlo simulations of the HP model \\ (the ``Ising model'' of protein folding)}

\author[1]{Ying Wai Li}
\author[2]{Thomas W\"ust}
\author[1]{David P. Landau}

\address[1]{Center for Simulational Physics, The University of Georgia, Athens, GA 30602, U.S.A.}
\address[2]{Swiss Federal Research Institute WSL, CH-8903 Birmensdorf, Switzerland}

\begin{abstract}

Using Wang-Landau sampling with suitable Monte Carlo trial moves (pull moves 
and bond-rebridging moves combined) we have determined the density of states 
and thermodynamic properties for a short sequence of the HP protein model. 
For free chains these proteins are known to first undergo a collapse 
``transition'' to a globule state followed by a second ``transition'' into a 
native state. When placed in the proximity of an attractive surface, there 
is a competition between surface adsorption and folding that leads to an 
intriguing sequence of ``transitions''. These transitions depend upon the 
relative interaction strengths and are largely inaccessible to ``standard'' 
Monte Carlo methods.

\end{abstract}


\end{frontmatter}


\section{Introduction}
\label{intro}

Over the last few decades, considerable effort has been devoted to the protein folding 
problem and yet major challenges remain. The complexity arises from enormous 
combinatorics of the 20 amino acids, forming multitudinous possible protein sequences 
to which the structure is closely related, and hence its biological function \cite{branden}. 
Complicated interactions among residues also result in rough free energy landscapes 
\cite{bryngelson}. Nevertheless, the underlying principles for protein structure prediction 
from the sequence are still unclear \cite{dill}. On the other hand, technical limitations 
also hinder experimental progress in obtaining protein structures using x-ray diffraction 
or NMR. Computer simulation can shed light on the problem where both theoretical and 
experimental studies are facing bottlenecks. Similar to the importance of the Ising 
model in the study of magnetism, simple protein models that embrace merely essential 
features are amenable to attack by a wide range of numerical methods, including 
computer simulations. This also allows understanding the folding problem from a 
macroscopic perspective without distraction from unnecessary details \cite{banavar}.


\section{Model and Methods}
\label{model}

The HP model is a simple, prototypical lattice protein model. It consists of 
only two types of monomers, hydrophobic (H) and polar (P), in a sequence chosen 
to mimic a real protein \cite{dill2}. Interactions are restricted to an attractive 
coupling, $\epsilon_{HH}$, between non-bonded hydrophobic monomers occupying 
nearest-neighbor sites. These are introduced to capture the hydrophobic effect, 
which is considered as the ``driving force'' of protein folding in forming tertiary 
structures. Despite its simplicity, the HP model shows unexpectedly rich 
thermodynamic behaviors and is challenging to study. Finding the ground state of an 
HP sequence is itself an NP-hard problem \cite{berger, crescenzi}.

It should be pointed out that since the HP chain is constructed specifically to 
represent an individual protein, thermodynamic and structural properties of the 
chain depend on its chain length and the sequence of H and P monomers uniquely. 
In this case, there does not exist finite size scaling for this model, and thus 
there are no real phases and transitions in the thermodynamic limit.

To understand how a protein interacts with an attractive substrate, a model 
surface which attracts both types of monomers with strength $\epsilon_s$ is 
introduced to the 3-dimensional lattice as an $xy$-plane placed at $z = 0$. 
A second, non-attractive wall is placed at $z = h_w$ to bound the HP chain 
from above. Periodic boundaries are used for the $x$ and $y$ directions. The 
Hamiltonian of the model is then represented as:
\begin{equation}
\mathcal{H} = - \epsilon_{HH} n_{HH} - \epsilon_s n_s,
\label{eq:hamiltonian}
\end{equation}

\noindent
$n_{HH}$ being the number of H-H interacting pairs and $n_s$ being the number 
of monomers adjacent to the bottom surface. Such a system has first been 
introduced and studied in Ref. \cite{bachmann}.

We have implemented two inventive Monte Carlo trial moves, namely, pull moves 
\cite{lesh} and bond-rebridging moves \cite{deutsch}, which are efficient both 
in compact and elongated configurations \cite{wuest}. Employing Wang-Landau 
sampling \cite{wang} we have obtained the density of states in energy $g(E)$, 
from which the partition function $Z$ can be calculated:
\begin{equation}
Z = \sum_{i}{g(E_i)e^{-E_i/kT}},
\label{eq:partitionFunction}
\end{equation}

\noindent
where $k$ is the Boltzmann constant, $T$ is temperature and the sum runs over 
all possible energies. The average energy $\left\langle E \right\rangle$ then 
follows:
\begin{equation}
\left\langle E \right\rangle = \frac{1}{Z}\sum_{i}{E_ig(E_i)e^{-E_i/kT}},
\label{eq:energy}
\end{equation}

\noindent
and the specific heat can be calculated from the fluctuation of energy:
\begin{equation}
C_v = \frac{1}{kT^2}(\left\langle E^2 \right\rangle - \left\langle E \right\rangle^2).
\label{eq:specificHeat}
\end{equation}

In Wang-Landau sampling, $g(E)$ is estimated iteratively. During the simulation, 
a multiplicative modification factor $f$ is used to update $g(E)$ with an 
initial value $f_{init} = e^1$. A histogram in energy $H(E)$ is accumulated 
besides $g(E)$. When a ``flat'' histogram is attained, the simulation is brought 
to the next iteration: $H(E)$ is reset and $f$ is reduced, $f_{new} = \sqrt{f}$. 
A histogram is said to be ``flat'' when all entries in $H(E)$ are greater than 
$p \times H_{ave}$, where $p$ is the flatness criterion and $H_{ave}$ is the 
average of all entries in $H(E)$. To yield reliable estimates for $g(E)$, all 
results presented in this work are obtained by using a flatness criterion 
$p = 0.8$. The simulation stops when the natural log of the final modification 
factor, $ln(f_{final})$, reaches a preset minimum value of $10^{-8}$.


\section{Results}
\label{results}

\begin{figure}[h]
  \centering
  \includegraphics[width=0.4\columnwidth]{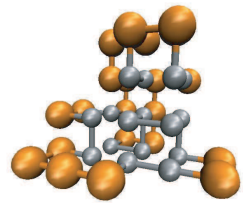}
  \caption{(Color online.) Ground state of the 36mer in free space ($E = -18$, $n_{HH} = 18$). 
           The HP chain is displayed with larger polar and smaller hydrophobic residues.}
  \label{fig:3Dfree}
\end{figure}

As a first step, we have simulated an HP sequence with 36 beads 
(PPPHHPPHHPPPPPHHHHHHHPPHHPPPPHHPPHPP), which has first been studied in Ref. \cite{swetnam}. 
In free space, the HP chain undergoes a coil-globule transition as temperature 
decreases, resulting in a compact hydrophobic core with polar residues residing on the 
exterior to screen the core from the polar solvents. See Figure \ref{fig:3Dfree}. 
Similar behaviors have also been observed for other lengths and sequences 
\cite{bachmann2, wuest2, li}.

With a surface that attracts both H and P monomers, the HP chain exhibits a much 
richer ``transition'' hierarchy due to the competition between surface adsorption 
and attraction within the polymer. To illustrate the effects of the surface 
strength on the ``transition'' behaviors, both a weak surface ($\epsilon_{HH} = 12, 
\epsilon_s = 1$) and a strong surface ($\epsilon_{HH} = 1, \epsilon_s = 1$) are 
considered. 15 independent runs were performed to obtain statistical errors in 
each case. CPU time grows with the size of the energy range of the systems: it 
takes about 15 minutes to finish a simulation on an AMD Opteron dual-core 2.2 GHz 
processor for the surface-free case using 19 energy bins, 5-10 hours for the strong 
attractive surface case with 51 energy bins, but more than 10 days are generally 
needed for the weak attractive surface where there are 242 energy bins in the full 
energy range.

\begin{figure}[ht]
  \includegraphics[width=1.0\columnwidth]{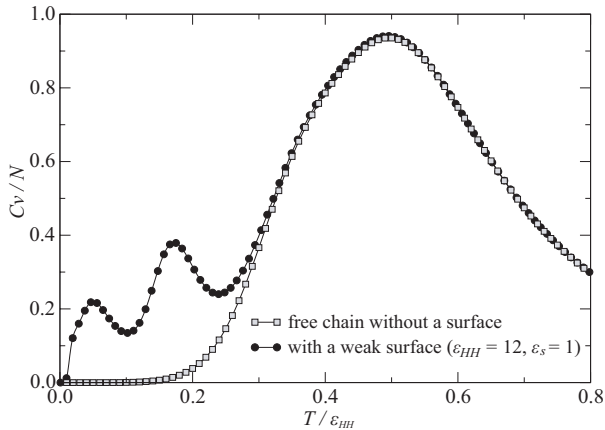}
  \caption{Specific heat of a HP 36mer: (i) in a 3D free space, (ii) interacting with 
           a weakly attractive surface ($\epsilon_{HH} = 12, \epsilon_s = 1$, $h_w = 
           37$). Error bars are smaller than the data points.}
  \label{fig:Cv_compare}
\end{figure}

Figure \ref{fig:Cv_compare} shows a typical specific heat of an HP chain in 
three-dimensional free space, as well as that when it interacts with a weakly 
attractive surface. In the latter case the height of the non-attractive wall is 
set to be $h_w = N + 1 = 37$, i.e., there are 36 layers between the two horizontal 
surfaces, the 36mer can touch both surfaces with its ends only when it is a fully 
stretched, vertical chain. While there is only a single peak corresponding 
to the collapse ``transition'' at $T / \epsilon_{HH} \approx 0.5$ for a free chain, 
three peaks are observed in the case with a weakly attractive surface. From this 
comparison it is obvious that the two peaks in the low temperature regime are due 
to the influence of the attractive substrate.

\begin{figure*}[t]
  \centering
  \includegraphics[width=2.0\columnwidth]{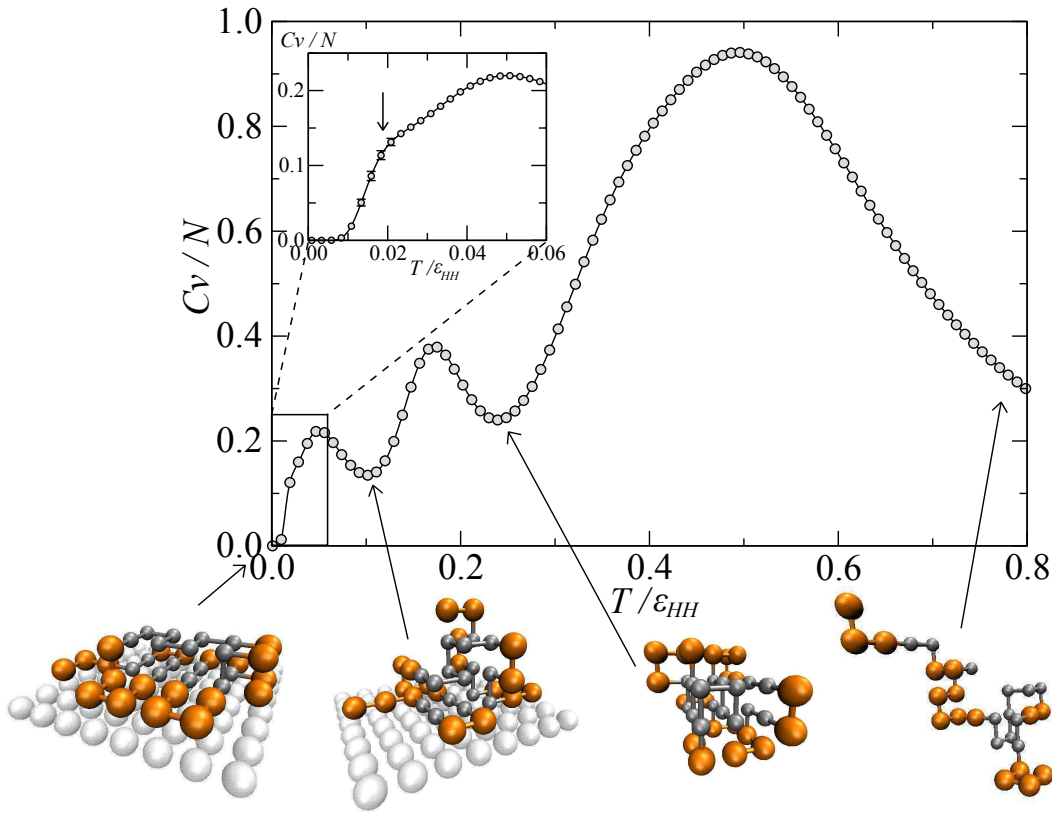}
  \caption{(Color online.) Specific heat of an HP 36mer interacting with a weak 
           attractive surface ($\epsilon_{HH} = 12, \epsilon_s = 1$), $h_w = 37$. 
           Error bars smaller than the data points are not shown. Typical structures 
           are displayed with larger polar and smaller hydrophobic residues.}
  \label{fig:Cv_weaksurface}
\end{figure*}

The largest peak which overlaps with the free chain's specific heat represents the same 
coil-globule transition: during this stage, the HP polymer transforms from an extended 
coil-like structure to a desorbed, but compact, globule. Typical resulting structures 
are principally the same as the ground states of the free chain in the absence of the 
surface, as shown in Figure \ref{fig:Cv_weaksurface}.

The middle peak at $T / \epsilon_{HH} \approx 0.18$ signals adsorption 
during which the compact HP globule ``docks'' at the surface with the hydrophobic core 
remaining intact. The conformation spans several layers vertically and the total energy 
of the system is lowered slightly due to contact with the surface, by an amount dependent 
upon the number of surface contacts.

Further decrease in temperature brings the system to the third transition at 
$T  / \epsilon_{HH} \approx 0.05$ where it strives for a maximum number of surface 
interactions without sacrificing an intact, energetically minimized hydrophobic 
core. With a large value of $\epsilon_{HH}$, forming H-H contacts is immensely more 
energetically favorable than forming surface contacts.

\begin{figure}[ht]
  \includegraphics[width=0.48\columnwidth]{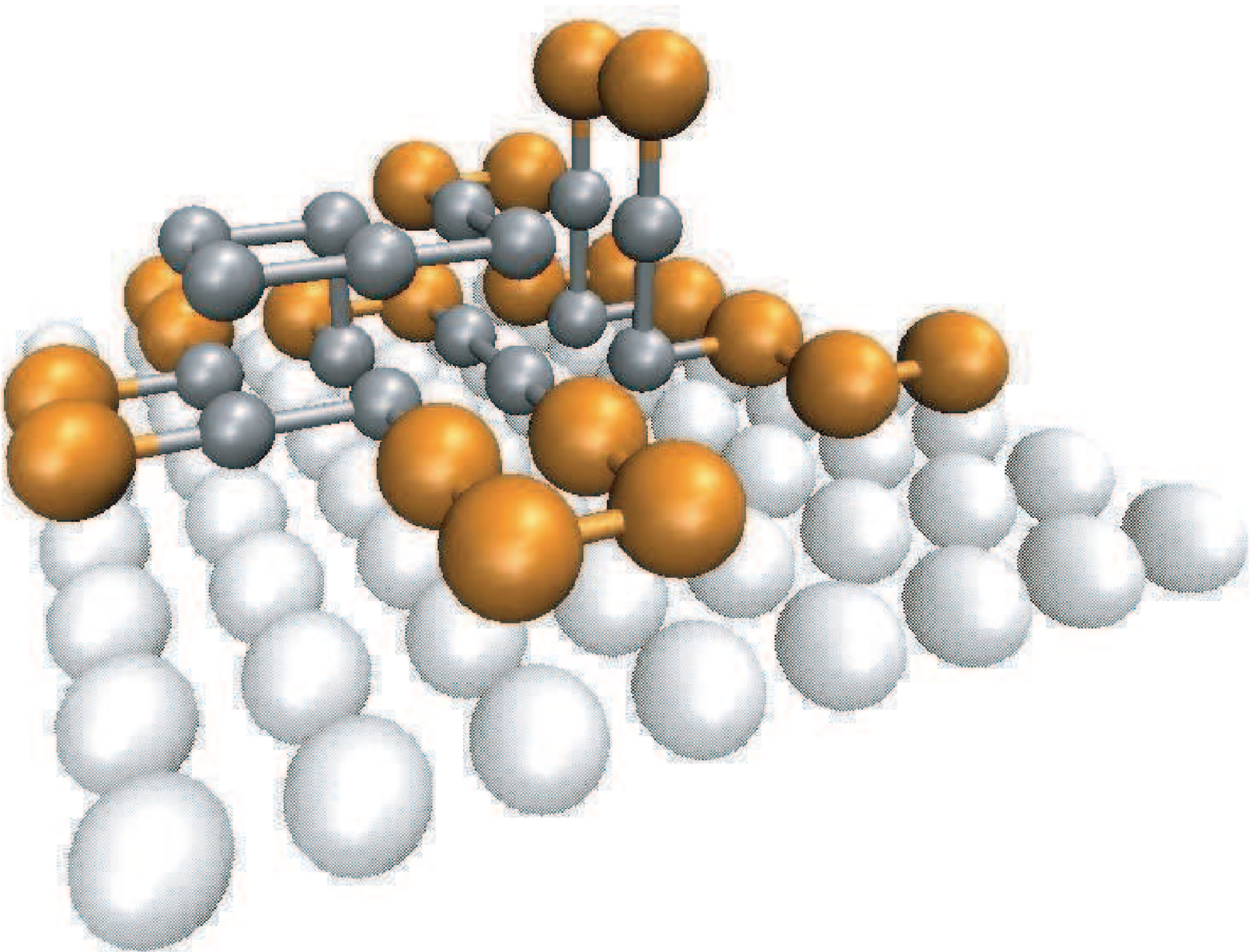}
  \includegraphics[width=0.46\columnwidth]{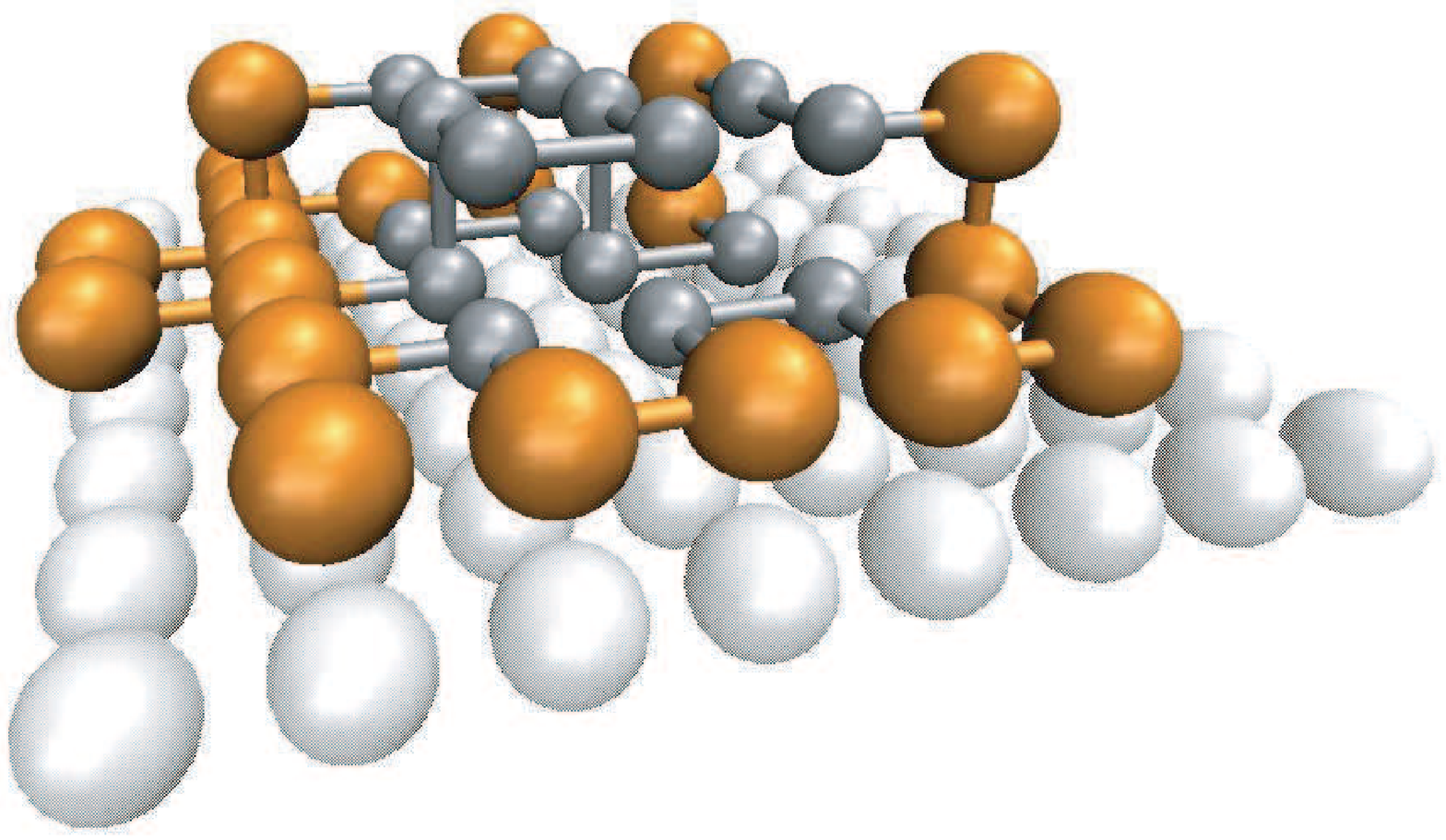}
  \caption{(Color online.) First excited states of the 36mer interacting with 
           a weak attractive surface, $E = -240$ $(n_{HH} = 18, n_s = 24)$. 
           The HP chains are displayed with larger polar and smaller 
           hydrophobic residues.}
  \label{fig:1st_excited_weak}
\end{figure}

A closer look at the low temperature regime reveals a subtle shoulder at 
$T / \epsilon_{HH} \approx 0.02$, which has not been found in Ref. 
\cite{swetnam}. This shoulder is a crossover signal owing to a transition from 
the ground state to the first few excited states at low temperature, that is 
similar to the case reported in the investigation of freezing and collapse of 
homopolymers \cite{wuest, vogel}. In our system, this excitation is an effect 
purely due to the existence of the surface, which is supported by the analyses 
on other structural parameters (not presented here) and also by observing the 
typical states at $T / \epsilon_{HH} = 0$ and $T / \epsilon_{HH} \approx 0.02$. 
At $T / \epsilon_{HH} = 0$ where the polymer occupies just ground states, only 
rectangular cores are able to maximize the number of surface contacts. This 
gives the ground state energy $E = -241$ $(n_{HH} = 18, n_s = 25)$. An example 
of such a structure is shown in the bottom left-hand corner of Figure 
\ref{fig:Cv_weaksurface}. At $T / \epsilon_{HH} \approx 0.02$ where first and 
second excited states dominate, two core shapes are observed: rectangular and 
``L-shape'' as shown in Figure \ref{fig:1st_excited_weak}. Since these two 
cores have the same energetic contribution ($n_{HH} = 18$, with 8 H monomers 
interacting with the surface for both cases), the excitation from the ground 
state to the first few excited states can only be due to the decrease in surface 
contacts of the P monomers.

\begin{figure*}[t]
  \centering  
  \includegraphics[width=2.0\columnwidth]{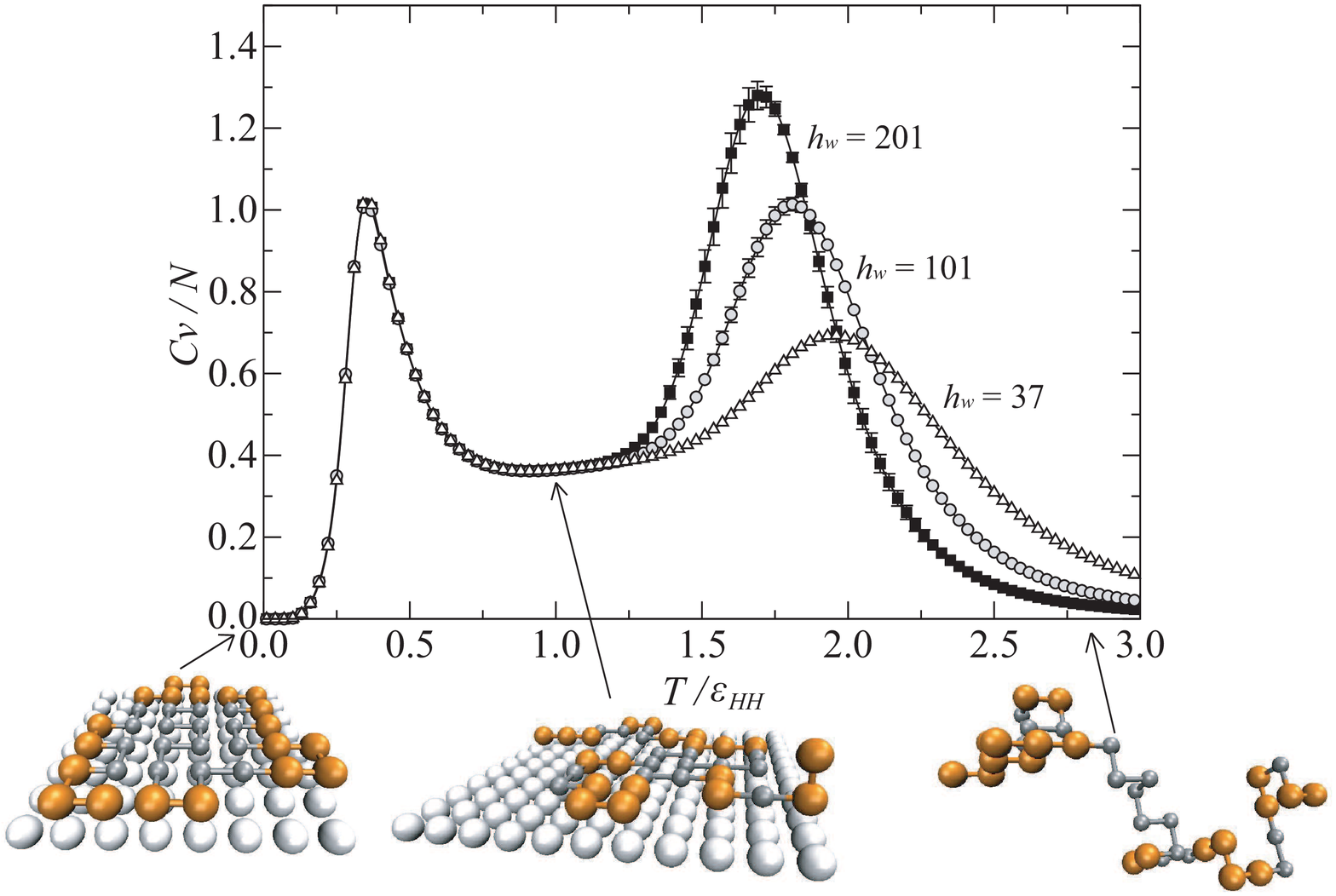}
  \caption{(Color online.) Specific heat of an HP 36mer interacting with a strong 
           attractive surface ($\epsilon_{HH} = 1, \epsilon_s = 1$). Error bars 
           smaller than the data points are not shown. Typical structures are 
           displayed with larger polar and smaller hydrophobic residues.}
  \label{fig:Cv_entropic}
\end{figure*}

On the other hand, when the HP chain is brought near a strong surface field, 
the transition behavior is quite different from the previous case. We observe 
only two peaks in the specific heat, see Figure \ref{fig:Cv_entropic}.
The one at higher temperature corresponds to an adsorption transition, and the 
other one, at $T / \epsilon_{HH} \approx 0.3$, corresponds to the collapse 
transition that takes place on the surface. At high temperature, the extended 
HP chain first attaches to the surface until most monomers touch it. This 
adsorbed, yet expanded structure then undergoes a second transition to 
maximize the number of H-H contacts and finally achieves a film-like, 
two-dimensional compact structure.

\begin{figure}[ht]
  \includegraphics[width=0.48\columnwidth]{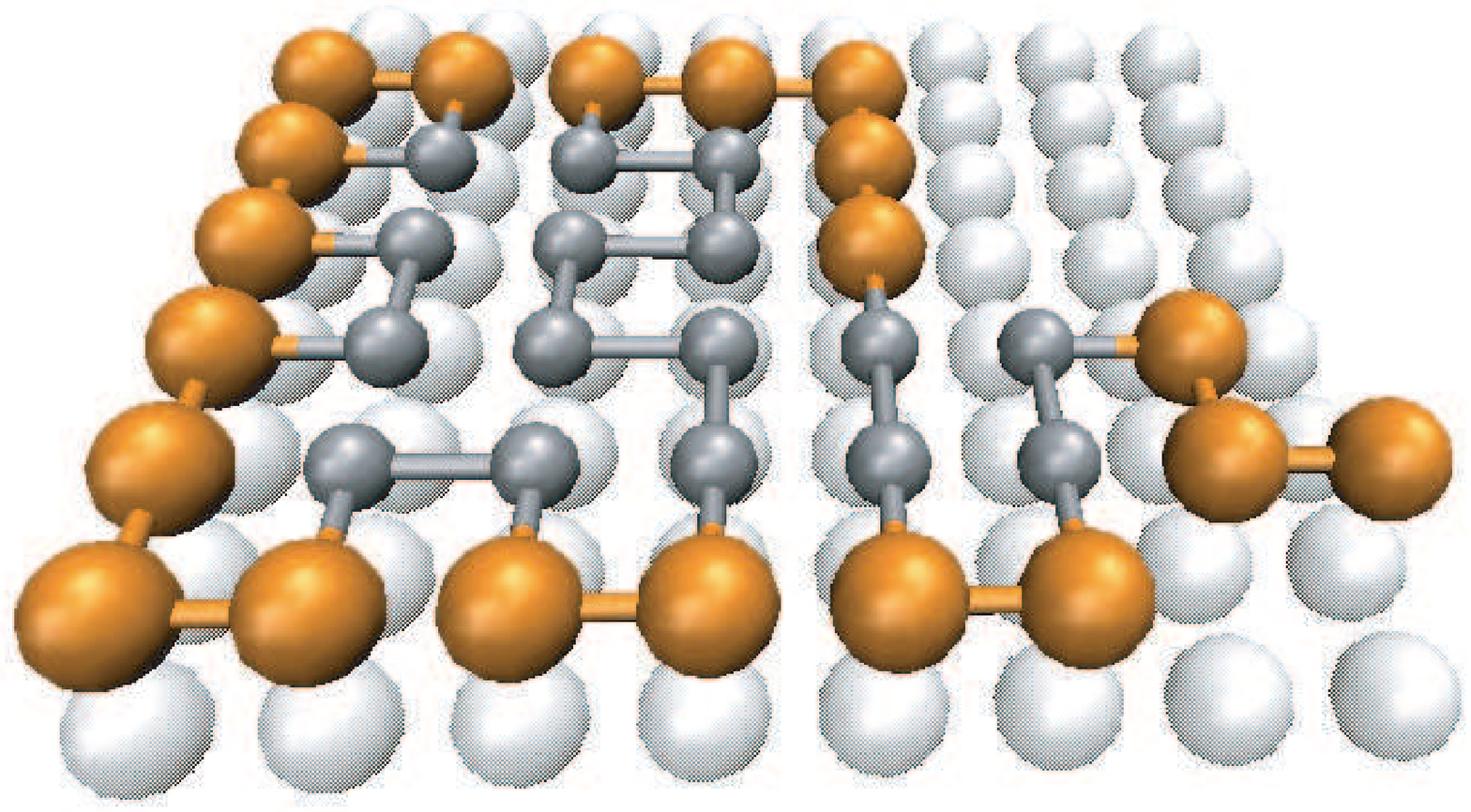}
  \includegraphics[width=0.48\columnwidth]{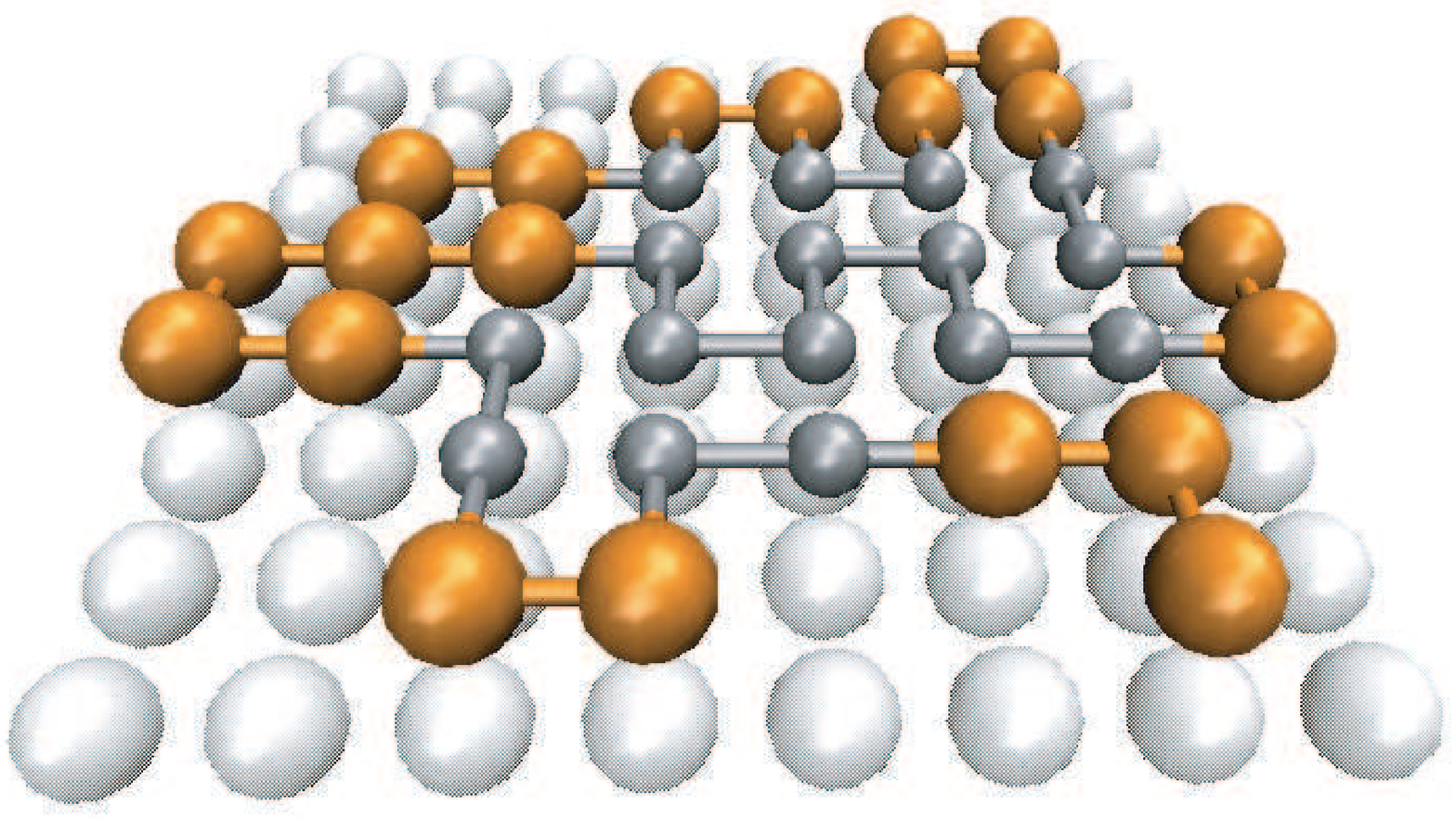}
  \caption{(Color online.) First excited states of the 36mer interacting 
           with a strong attractive surface, $E = -49$ $(n_{HH} = 13, 
           n_s = 36)$. The HP chains are displayed with larger polar and 
           smaller hydrophobic residues.}
  \label{fig:1st_excited_strong}
\end{figure}

Unlike the previous case with a weakly attractive surface, there is no 
shoulder in the very low temperature regime. This is because the excitation 
from the ground state ($E = -50, n_{HH} = 14, n_s = 36$, as shown in the 
bottom left-hand corner of Figure \ref{fig:Cv_entropic}) to the first 
excited states ($E = -49, n_{HH} = 13, n_s = 36$) in this case is mainly 
achieved by the destruction of the H-H contacts, while the entire chain 
still remains completely attached to the substrate. Figure 
\ref{fig:1st_excited_strong} shows two examples of first excited states 
having different hydrophobic cores from that of the ground state. Therefore, 
the effect of the excitation cannot be singled out to give an individual 
signal, but is rather included within the collapse transition peak.

Figure \ref{fig:Cv_entropic} also shows the influence of the entropic effects 
on the transition behavior due to the height $h_w$ of the non-interacting wall 
bounding the chain from above. Obviously the low-temperature collapse is not 
affected at all, as it takes place on the attractive surface which is not 
related to the steric upper surface. The desorption-adsorption transition, 
however, shows systematic dependence on $h_w$. A smaller $h_w$ restricts the 
vertical movements of the HP chain to a larger extent, resulting in a smaller 
entropy gain (and thus a less pronounced peak in the specific heat) as less 
translational variations of the same configuration are allowed. The chain is 
also more likely to be on the attractive surface, resulting in a higher 
adsorption temperature. This dependence of the adsorption transition peak on 
$h_w$ is also reported recently \cite{moeddel}, where an off-lattice 
homopolymer model is used to study polymer adsorption.


\section{Summary and Outlook}
\label{summary}

To summarize, we studied protein folding using Wang-Landau sampling with a 
minimalistic lattice model, the HP model. Two Monte Carlo trial moves, pull moves 
and bond-rebridging moves, which work particularly well with our algorithm, were 
implemented. In this work, we have focused on investigating protein adsorption. 
With a weak surface, three-stage folding is identified, while two-stage folding 
is found with a strong surface. Varying the height of the upper non-interacting 
surface regulates the total entropy of the system, which affects the higher 
temperature ``transition''. Further work in progress includes longer benchmark 
HP sequences and the investigation of the thermodynamics of different structural 
quantities and their relationship with different ``transitions''.


\section{Acknowledgment}
\label{acknowledgment}

We would like to thank M. Bachmann, M. Laradji and T. Vogel for constructive 
discussions. This work is supported by NSF Grant DMR-0810223 and NIH Grant 
IR01GM075331. 


\end{document}